\documentclass[%
 reprint,
amsmath,amssymb,
aps,
]{revtex4-2}
\usepackage{arydshln}
\usepackage{multirow}

\usepackage{graphicx}% Include figure files

\usepackage{dcolumn}% Align table columns on decimal point
\usepackage{bm}% bold math

\usepackage[colorlinks,linkcolor=blue,citecolor=blue,urlcolor=blue]{hyperref}

\usepackage{booktabs}

\begin{document}

\preprint{}

\title{Thermal Hall conductivity of semi-metallic 
graphite dominated by ambipolar phonon drag}

\author{Qiaochao Xiang$^{1}$, Xiaokang Li$^{1,*}$, Xiaodong Guo$^{1}$, Zengwei Zhu$^{1,*}$ and Kamran Behnia$^{2,*}$}

\affiliation{
(1) Wuhan National High Magnetic Field Center, School of Physics, Huazhong University of Science and Technology, 430074 Wuhan, China\\ 
(2) Laboratoire de Physique et d'\'Etude des Mat\'eriaux \\ (ESPCI - CNRS - Sorbonne Universit\'e), PSL Research University, 75005 Paris, France\\
}
\date{\today}

\begin{abstract}
It is now known that in addition to electrons, other quasi-particles such as phonons and magnons can also generate a thermal Hall signal. Graphite is a semimetal with extremely mobile charge carriers of both signs and a large lattice thermal conductivity. We present a study of the thermal Hall effect in highly oriented pyrolytic graphite (HOPG) samples with electronic, phononic and phonon drag contributions to the thermal Hall signal. The measured thermal Hall conductivity ($\kappa_{xy}$) is two orders of magnitude higher than what is expected by electronic carriers according to the electrical Hall conductivity and the Wiedemann-Franz law, yielding a record Hall Lorenz number of $164.9\times10^{-8}V^2 K^{-2}$ ($\sim$67$L_0$) - the largest ever observed in a metal. The temperature dependence of the thermal Hall conductivity significantly differs from its longitudinal counterpart, ruling out a purely phononic origin of the non-electronic component. Based on the temperature dependence and the amplitudes of the Seebeck and Nernst responses, we demonstrate that ambipolar phonon drag dominates the thermal Hall response of graphite.

\end{abstract}

\maketitle

The thermal Hall effect, the thermal analog of the electric Hall effect, has been recently explored in numerous quantum materials~\cite{Strohm2005,Onose2008,Onose2010,Hirsch2015,Watanabe2016,Ideue2017,Sugii2017,Li2017,Kasahara2018,Grissonnanche2019,Li2020,Grissonnanche2020,Boulanger2020,Akazawa2020,Yamashita2020,Sim2021,Chen2022,Uehara2022,Jiang2022,Bruin2022,Li2023,Chen2024,Chen2024-2,Ataei2024,Meng2024,Li2025}. The list includes cases where the generators of the signal are identified as electrons~\cite{Zhang2000,Onose2008,Li2017}, magnons~\cite{Onose2010} or phonons ~\cite{Strohm2005,Sugii2017,Li2020,Grissonnanche2020,Boulanger2020,Sim2021,Chen2022,Uehara2022,Li2023,Chen2024,Chen2024-2,Ataei2024,Li2025}. In some cases~\cite{Hirsch2015,Watanabe2016,Kasahara2018,Grissonnanche2019,Yamashita2020}, other neutral quasi-particles have been invoked. In other cases, the thermal Hall signal has been interpreted in a framework of interplay between different species of heat carriers~\cite{Ideue2017,Akazawa2020,Jiang2022,Bruin2022,Meng2024}.

\begin{figure*}[ht]
\centering
\includegraphics[width=1.0\linewidth]{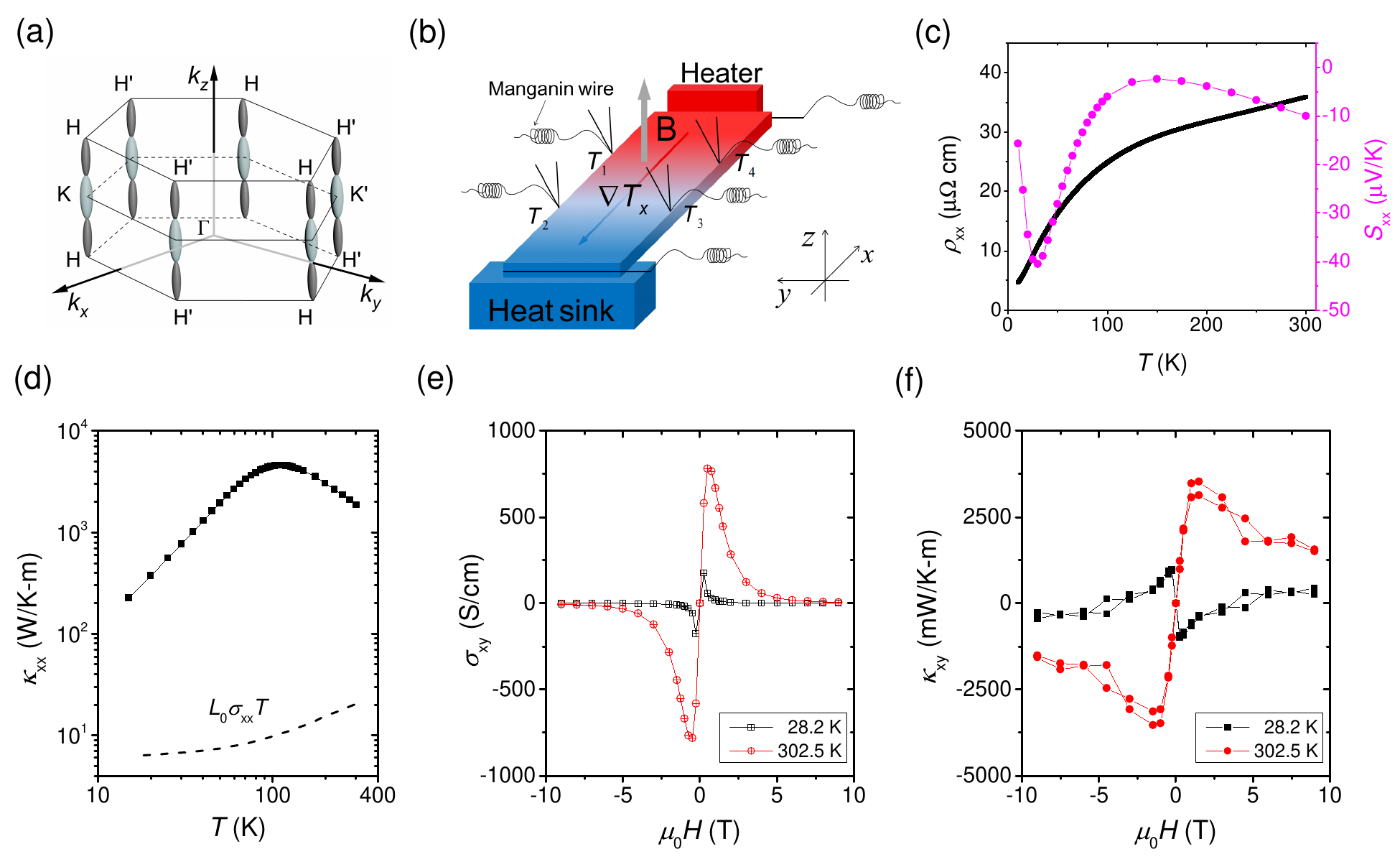} 
\caption{\textbf{Brillouin zone, experimental setup, longitudinal transport properties, and electric/thermal Hall conductivity in graphite.} (a) The Brillouin zone, electron and hole Fermi pockets in graphite. (b) Experimental configuration for measurement of longitudinal/transverse electric, thermoelectric, and thermal transport. (c) Temperature-dependent longitudinal resistivity ($\rho_{xx}$) and Seebeck coefficient ($S_{xx}$), showing a phonon drag peak near 30 K in $S_{xx}$. (d) Temperature dependence of longitudinal thermal conductivity ($\kappa_{xx}$). The dashed curve indicates negligible electronic contribution ($\kappa_{xx}^e$) estimated via $L_0\sigma_{xx}T$. (e) Field dependence of electric Hall conductivity ($\sigma_{xy}$) at 28.2 K (low temperature) and 302.5 K (room temperature), exhibiting sharp peaks. (f) Field-dependent thermal Hall conductivity ($\kappa_{xy}$) at 28.2 K and 302.5 K, showing gentler peaks with sign reversal between temperatures. The field is swept from positive to negative and then from negative to positive to ensure signal stability. Additional data points are provided in Supplementary Material~\cite{SM}.
}
\label{fig:THE}
\end{figure*}

Graphite, a Bernal stack of graphene layers, has been explored for decades \cite{brandt1988semimetals}. It is a compensated semimetal with high-mobility holes and electrons of equal density (Figure~\ref{fig:THE}a) and a very large phonon thermal conductivity~\cite{Klein1961,Chung2002,klemens1994,machida2020}. Extensive studies have documented its transport properties, such as longitudinal and Hall resistivity, Seebeck and Nernst coefficients and thermal conductivity~\cite{Soule1958,zhu2010,klemens1994,machida2020,wang2020,ye2024}. However, its transverse Hall thermal conductivity has never been explored before. Graphite hosts charge carriers of both signs as well as acoustic phonons with record velocity. Therefore, its study can provide a very useful input to the ongoing quest to understand what generates a thermal Hall effect beyond what is expected by the Wiedemann-Franz law.

In this study, we report measurements of the thermal Hall effect in highly oriented pyrolytic graphite (HOPG), revealing three key findings. First, we observe a large negative thermal Hall conductivity: $\kappa_{xy}$ becomes as large as -818.9 mW/(K·m) at 28.2 K under 0.25 T. It exceeds the electronic contribution predicted by the Wiedemann-Franz law ($L_0 \sigma_{xy} T$) by two orders of magnitude, leading to a record Hall Lorenz number of $164.9\times10^{-8}V^2 K^{-2}$ ($\sim$67$L_0$) - the largest reported in any metal. This result stands in contrast to other metallic systems where the Wiedemann-Franz law holds approximately~\cite{Arenz1982, Li2017}. This substantial discrepancy clearly indicates a contribution by quasi-particles other than electrons. Second, we find that $\kappa_{xy}$ reverses sign near 100 K. At room temperature and in a magnetic field of 1 T, its magnitude attains 3300 mW/(K·m). In contrast to what has been observed in a wide variety of insulators \cite{Li2023, Behnia2025}, the thermal Hall conductivity, $\kappa_{xy}$, and the longitudinal thermal conductivity, $\kappa_{xx}$, do not peak at the same temperature and the former changes sign. This rules out a purely phononic origin. Third, through systematic comparison of Nernst ($\alpha_{xy}$) and thermal Hall ($\kappa_{xy}$) conductivities, we identify ambipolar phonon drag - a hydrodynamic interplay between electrons, holes, and phonons - as the most plausible origin of this additional contribution. This interpretation is further corroborated by our quantification of the phonon drag Seebeck coefficient. Our results demonstrate that the combination of highly conductive phonons and high-mobility charge carriers of both signs can produce an exceptionally large thermal Hall response, surpassing the limits set by the Wiedemann–Franz law.

\begin{figure}[ht]
\centering
\includegraphics[width=1\linewidth]{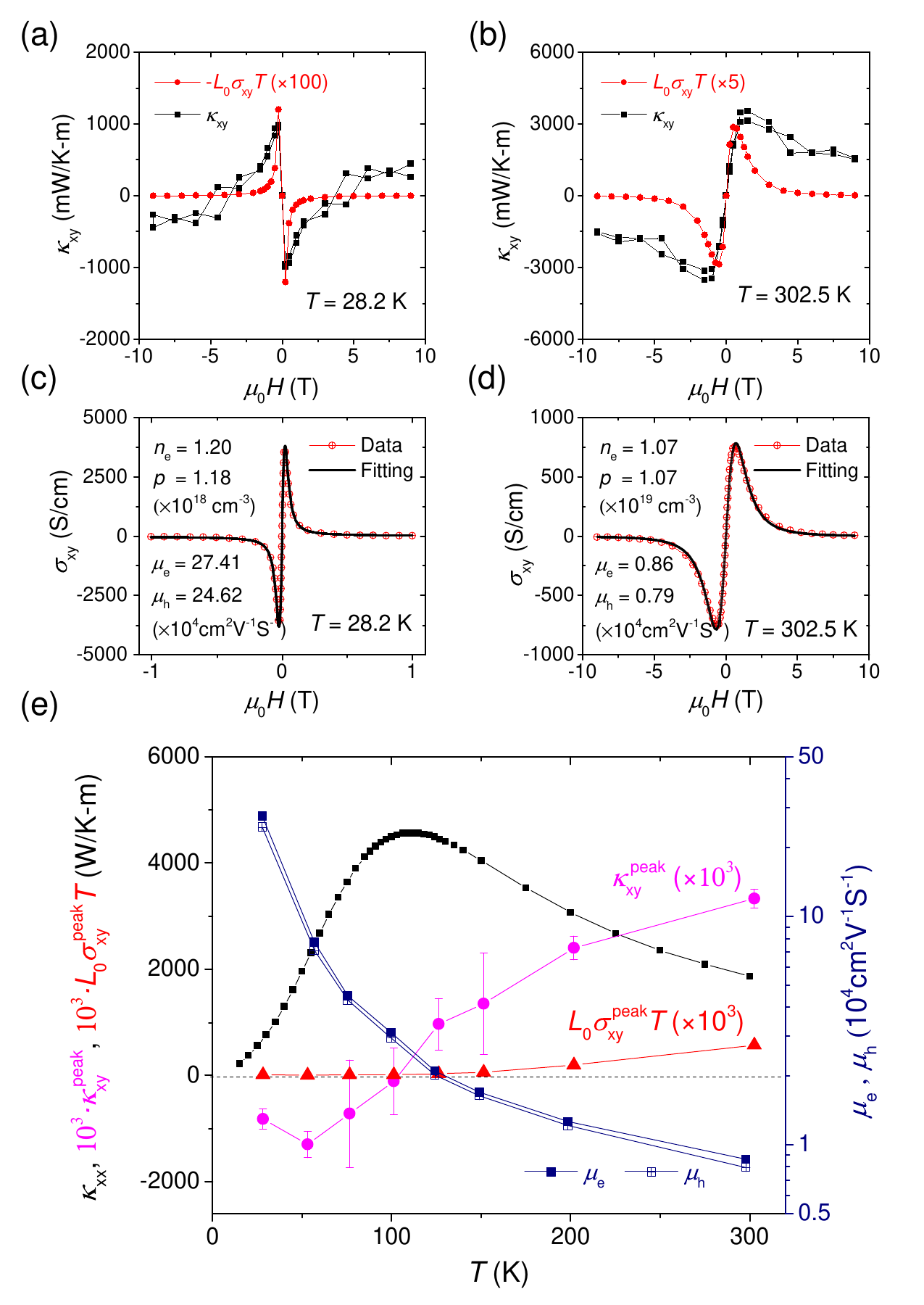} 
\caption{\textbf{Comparison of $\kappa_{xx}$, $\kappa_{xy}$ and $L_0\sigma_{xy}T$.} (a-b) Measured thermal Hall conductivity $\kappa_{xy}$ versus the theoretically expected electronic contribution $\kappa_{xy}^e = L_0\sigma_{xy}T$. The experimental values exceed the theoretical predictions by two orders of magnitude at 28.2 K and by a factor of five at 302.5 K, indicating a significant enhancement mechanism beyond electronic contributions. (c-d)  Hall conductivity curves fit to a two-band model ($\sigma_H(B) = \frac{ne\mu_e^2B}{1+\mu_e^2B^2} - \frac{pe\mu_h^2B}{1+\mu_h^2B^2}$). (e) Temperature-dependence of $\kappa_{xx}$, $\kappa_{xy}$ and $L_0\sigma_{xy}T$. Also shown are  carrier mobilities ($\mu_e$ and $\mu_h$). The contrasting trends between $\kappa_{xy}$ and $\kappa_{xx}$ suggest distinct physical origins for these transport phenomena. }
\label{fig:Comparison}
\end{figure}

\begin{figure*}[ht]
\centering
\includegraphics[width=1.0\linewidth]{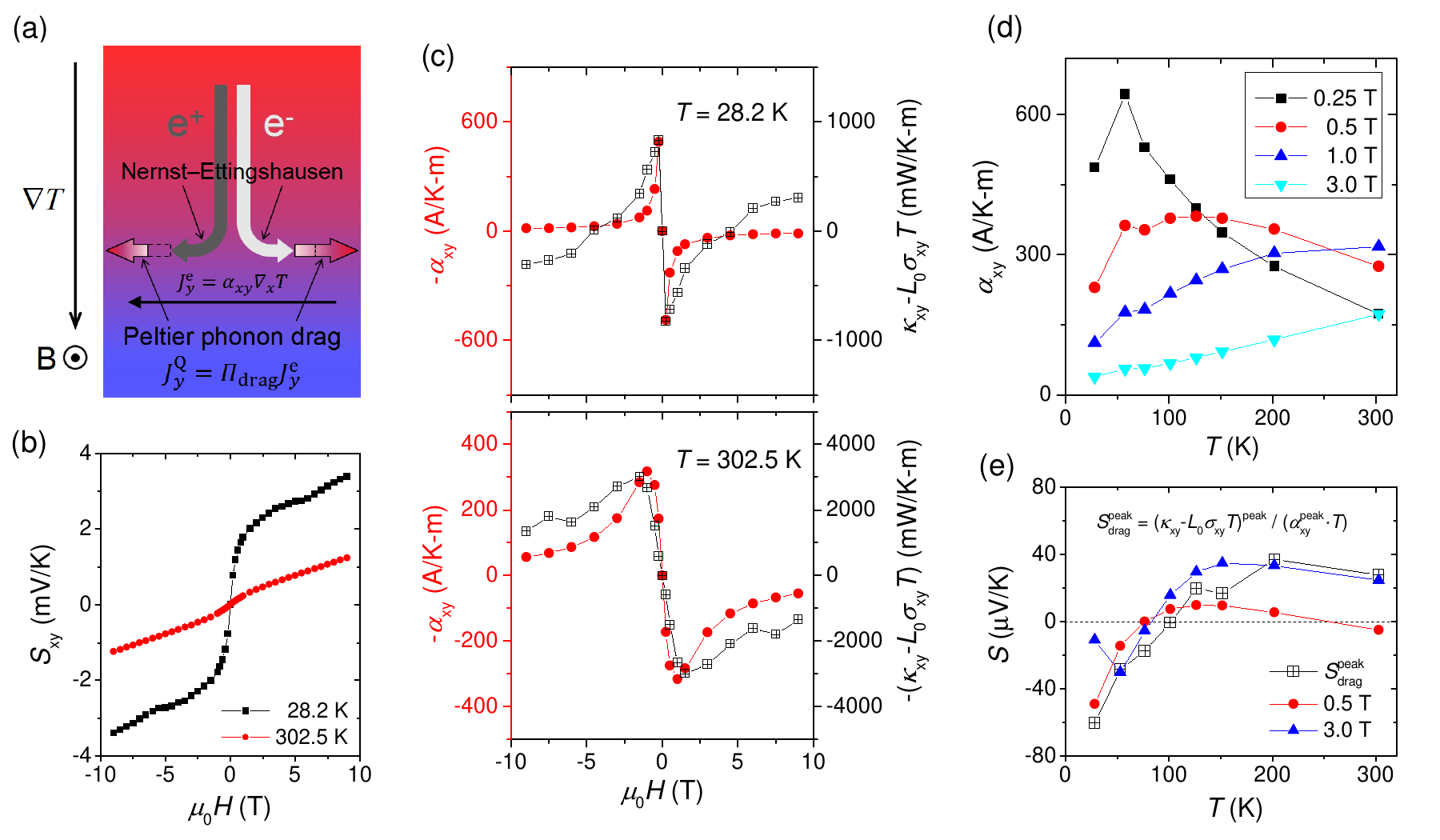} 
\caption{\textbf{Transverse ambipolar phonon drag.} (a) 
%Schematic of transverse electric (top) and thermal (bottom) diffusion processes. An electric field drives carrier motion: electrons and holes flow in opposite directions, but Lorentz force deflects both to the same side, canceling transverse electrical signals. Thermal gradient drives motion similarly, with Lorentz force deflecting carriers to opposite sides, canceling transverse thermal signals. (b) 
Schematic of thermal Hall phonon drag in compensated systems. Momentum-conserving collisions between carriers and phonons generate a transverse temperature gradient. The transverse electronic current ($J^e_y = \alpha_{xy} \nabla T_x$) induces transverse phonon momentum flow, resulting in phonon heat flow ($J^Q_y = \Pi_{drag} J^e_y$). (b) Field dependence of Nernst signal ($S_{xy}$). (c-d) Field and temperature dependence of Nernst conductivity ($\alpha_{xy}$), extracted from $S_{xy}$, $S_{xx}$, $\rho_{xy}$, $\rho_{xx}$ (see Supplementary Material~\cite{SM} for details). (e) Extracted temperature-dependent phonon drag Seebeck coefficient $S_{drag}^{peak} = (\kappa_{xy} - L_0\sigma_{xy}T)^{peak} / (\alpha_{xy}^{peak} T)$, compared to the measured Seebeck signal under field of 0.5 T (low temperature) and 3 T (high temperature). At 28.2 K, it reaches -60 $\mu $V/K, consistent with the peak measured in Figure~\ref{fig:THE}c.}
\label{fig:Phonondrag}
\end{figure*}

The experimental setup for measuring longitudinal and transverse electric, thermoelectric, and thermal transport properties is shown in  Figure~\ref{fig:THE}b. Four symmetrical thermocouples are longitudinally positioned on both sides of the sample, enabling the collection and cross-verification of two sets of transverse temperature difference data ($T_2-T_3$ and $T_1-T_4$) to ensure signal reliability and homogeneity. Figure~\ref{fig:THE}c-d present the temperature-dependent longitudinal transport signals, including resistivity ($\rho_{xx}$), Seebeck coefficient ($S_{xx}$), and thermal conductivity ($\kappa_{xx}$). The Seebeck signal exhibits a phonon drag peak at approximately 30 K, consistent with what was reported previously~\cite{Takezawa1971}. The measured  $\kappa_{xx}$ exceeds the upper limit of electronic contribution $\kappa_{xx}^e$ (estimated by $L_0\sigma_{xx}T$) by two orders of magnitude, indicating that phonons dominate heat conduction in our temperature window of interest (from 20 K to 300 K).

Prior to thermal Hall effect measurements, we quantified the electric Hall response of our samples, which depends on the differential mobility of electrons and holes. As seen in Figure~\ref{fig:THE}e, the field-dependent Hall conductivity $\sigma_{xy}(B)$ at two different temperatures, exhibits a sharp peak below 1 T before gradually approaching zero at high magnetic field. Notably, $\sigma_{xy}$ is positive at both temperatures. This is in sharp contrast with the behavior of the thermal Hall conductivity shown in Figure~\ref{fig:THE}f: $\kappa_{xy}$ displays a temperature-dependent sign reversal (Figure~\ref{fig:Comparison}e). The negative $\kappa_{xy}$ at low temperatures becomes positive at room temperature. This sign reversal is accompanied by a broadening peak profile. As we will discuss subsequently, this phenomenon indicates  a change in the dominant momentum exchange mechanism between charge carriers and phonons.

Figure~\ref{fig:Comparison}a-b compares the measured thermal Hall conductivity $\kappa_{xy}$ with the electronic contribution $\kappa_{xy}^e = L_0\sigma_{xy}T$ predicted by the Wiedemann-Franz law. The experimental values exceed the theoretical prediction by two orders of magnitude at 28.2 K and by a factor of five at 302.5 K.  This pronounced deviation not only signals a major non-electronic contribution but also leads to a record-high Hall Lorenz number of $164.9\times10^{-8}V^2 K^{-2}$ ($\sim$67$L_0$) in any metal (see Table~\ref{tab:sample_properties}). Here, $L_0$ denotes the Sommerfeld value ($2.44\times10^{-8}V^2 K^{-2}$).

Figure~\ref{fig:Comparison}e (left $y$-axis) presents the temperature evolution of $\kappa_{xx}$, $\kappa_{xy}$ and $L_0\sigma_{xy}T$. In the case of the latter two quantities, each data point represents the peak in the  field-dependent curve at each temperature. Throughout the measured temperature range, the absolute value of $\kappa_{xy}$ exceeds by far $L_0\sigma_{xy}T$. Meanwhile, its temperature dependence is very different from the phonon-dominated $\kappa_{xx}$. This difference strongly suggests that phonons cannot be the unique (or even the most prominent) source of the large and sign-changing $\kappa_{xy}$. 

Figure~\ref{fig:Comparison}c-d show how the electric Hall conductivity can be fit to a two-band model at 28.2 K and 302.5 K, which are the extreme values of our temperature window. Carrier density as well as electron and hole mobilities can be extracted from these fits at each temperature. The temperature-induced change in the carrier density and mobility agrees with prior works~\cite{Klein1962,Soule1958, mcclure1963theory, spain1967electronic}, and the former suggests an effective band gap (0.02–0.04 eV) in graphite~\cite{Dresselhaus2002} (see Supplementary Material~\cite{SM} for details). As seen in Figure~\ref{fig:Comparison}e (right $y$-axis) electrons are slightly more mobile than holes in the whole temperature range. 

Phonon drag~\cite{Johnson1953,Herring1954,behnia2015fundamentals}, which refers to momentum exchange between charge carriers and phonons, is known to amplify the Seebeck effect of common semiconductors at low temperature.  Jiang \textit{et al.}~\cite{Jiang2022} proposed that it can also amplify the thermal Hall response when the Hall angle of charge carriers is very large but heat conduction is mostly phononic. The proposal, based on Herring's picture of phonon drag~\cite{Herring1954}, was applied to dilute metallic SrTiO$_3$~\cite{Jiang2022}. This conclusion was recently backed by a rigorous theoretical study employing the Kubo formula to quantify thermal and thermoelectric transport~\cite{Endo2024}.

Herring argued that the Peltier coefficient ($\Pi= ST$) is enhanced by the drag exerted on phonons by the electric current \cite{Herring1954}. Assuming an approximate proportionality between heat current and crystal momentum, he found that a phonon drag Peltier effect, $\Pi_{drag}$, of either sign can arise : 
\begin{equation}
    \Pi_{drag}=\pm \frac{m^*v_s^2}{e} f\frac{\tau_p}{\tau_e}
    \label{Herring}
\end{equation}

In this equation, $m^*$ is the effective mass, $v_s$ is the sound velocity, $e$ is the electron charge, and $\tau_p$ and $\tau_e$ are the phonon and electron scattering times, respectively. The parameter $0<f<1$ represents the fraction of phonon collisions transferring momentum to the electron bath. Thus, an electric current can generate a phonon energy flow.

Now, in the presence of a finite Nernst conductivity, $\alpha_{xy}$ links a longitudinal thermal gradient to a perpendicular charge current ($J^e_y = \alpha_{xy} \nabla T_x$) and this charge current generates a phonon drag Peltier response ($J^Q_y = \Pi_{drag} J^e_y$). This leads to : 

%Driven by the longitudinal thermal gradient, the transverse electronic charge flow can be expressed as:

%Where $\alpha_{xy}$ is the Nernst conductivity. \textcolor{red}{The ambipolar flow of quasiparticles under a thermal gradient can lead to an enhancement of the Nernst signal ($\alpha_{xy}$)~\cite{Bel2003} and the transverse charge flow {$J^e_y$}.} Momentum-conserving collisions between charge carriers and phonons will generate a transverse heat flow:
%\begin{equation}
	%	 = \Pi_{drag}\alpha_{xy} \nabla T_x
		\label{J^Q}
%\end{equation}
%Here, $\Pi_{drag}$ is the phonon drag Peltier coefficient. Using the Kelvin relation, the phonon drag component of Seebeck coefficient is $\Pi_{drag}/T$. Rearranging the equation~\ref{J^Q}, we get:

\begin{equation}
		\kappa_{xy}^{drag}= \Pi_{drag} \alpha_{xy} = S_{drag}T \alpha_{xy} 
		\label{kappadrag}
\end{equation}

 To check the relevance of Equation~\ref{kappadrag}, we need to quantify $\alpha_{xy} $. Figure~\ref{fig:Phonondrag}b presents the magnetic field dependence of the Nernst coefficient $S_{xy}$. It is noteworthy that the field dependence of $S_{xy}$ differs markedly from that of $\kappa_{xy}$. At 28.2 K, the hump around 6 T in the $S_{xy}$ curve is most plausibly due to a quantum oscillation in the Nernst response, detectable at this temperature~\cite{zhu2010}.

After measuring $S_{xy}$, $S_{xx}$, $\rho_{xy}$ and $\rho_{xx}$ (see Supplementary Material~\cite{SM} for details), we extracted the field-dependent Nernst conductivity $\alpha_{xy}(B)$ (Figure~\ref{fig:Phonondrag}c), whose field variation pattern is indeed strikingly similar to $\kappa_{xy}$, not only in its global shape  but also in the position of the peak. We also obtained the temperature-dependent Nernst conductivity $\alpha_{xy}(T)$ (Figure~\ref{fig:Phonondrag}d), which retains its sign across the whole temperature range. The observed consistency strongly supports our theoretical framework and provides a method to quantitatively determine the phonon drag component in Seebeck coefficients, as shown in Figure~\ref{fig:Phonondrag}e. The extracted phonon-drag Seebeck coefficient reaches -60 $\mu $V/K$ $ at 28.2 K, consistent with the measured phonon-drag peak in Figure~\ref{fig:THE}c, and is comparable to the measured Seebeck signal under field of 0.5 T (low temperature region) and 3 T (high temperature region), providing additional experimental verification for our model. Previous studies~\cite{Jay1970, Takezawa1971, Tsuzuku1972, Sugihara1972} have found that in graphite, the Seebeck and the Hall  coefficients vary from one sample to another, as a consequence of the presence of impurities and defects, which can dope the pristine semi-metal (See \cite{SM} for a comparison of our Seebeck data with previous studies). However, there is broad consensus that the Seebeck coefficient in graphite is dominated by phonon drag.
%There is a broad consensus, however, that the Seebeck coefficient is dominated by phonon drag in the whole temperature window of interest of our study. 
Moreover, Sugihara and co-workers~\cite{Sugihara1972} argued that phonon drag in graphite is strongly enhanced by a magnetic field and decreases with rising temperature. Our findings are consistent with this picture.

\begin{table}[htbp]
\centering
\scriptsize
\setlength{\tabcolsep}{2.5pt}
\caption{The Hall Lorenz number (${L}_{xy}$) in different metals, data from~\cite{Zhang2000,Matusiak2005, matusiak2006hall,Jiang2022,Shiomi2009,Onose2008,Li2017,xu2020finite,zhang2021topological,Zhang2024Abnormally}.}
\label{tab:sample_properties}
\begin{tabular}{lccccc}
\toprule
Sample & $\lvert\sigma_{xy}\rvert$ & $\lvert\kappa_{xy}$/$T$$\rvert$ & $T, B$ & $\lvert L_{xy}^{max}/L_0\rvert$ & Ref.\\
 & (S/cm) & (W\!/K$^2$\!-m) & (K, T) & - &-\\
\midrule
Cu & $-$ & $3.8\!\!\times\!\!10^{-2}$ & 350, 14 & 1.06 & \cite{Zhang2000}\\
YBa$_2$Cu$_3$O$_{6.95}$ & $-$ & $2.7\!\!\times\!\!10^{-5}$ & 320, 14 & 0.65 & \cite{Zhang2000}\\
EuBa$_2$Cu$_3$O$_{6.65}$ & $-$ & $-$ & 80, $-$ & 3.13 & \cite{Matusiak2005}\\
{La}$_{1.855}$Sr$_{0.145}$CuO$_4$ & $-$ & $-$ & 270, $-$ & 0.91 & \cite{matusiak2006hall}\\
SrTiO$_{3-\delta}$ & $1.7\!\!\times\!\!10^{2}$ & $5\!\!\times\!\!10^{-3}$ & 20, 12 & $12.0$ & \cite{Jiang2022}\\
Fe & $5\!\!\times\!\!10^{5}$ & 1.1 & 10, 9 & 0.91 & \cite{Shiomi2009}\\
Ni & $8.3\!\!\times\!\!10^{4}$ & $0.19$ & 10, 9 & 0.94 & \cite{Onose2008}\\
Mn$_3$Sn & 54 & $1.4\!\!\times\!\!10^{-4}$ & 300, $-$ & 1.08 & \cite{Li2017}\\
Mn$_3$Ge & $1.8\!\!\times\!\!10^{2}$ & $5.5\!\!\times\!\!10^{-4}$ & 5.6, $-$ & $1.23$ & \cite{xu2020finite}\\
Fe$_3$Sn$_2$ & $3.0\!\!\times\!\!10^{2}$ & $2.9\!\!\times\!\!10^{-4}$  & 300, 1.5 & 0.40 & \cite{zhang2021topological}\\
NdAlSi & $9.2\!\!\times\!\!10^{3}$ & $4.9\!\!\times\!\!10^{-2}$ & 20, 9 & 2.17 & \cite{Zhang2024Abnormally}\\
Graphite & $1.8\!\!\times\!\!10^{2}$ & $2.9\!\!\times\!\!10^{-2}$ & 28.2, 0.25 & 67.3& here\\
\bottomrule
\end{tabular}
\end{table}

Figure~\ref{fig:Phonondrag}a is a schematic diagram of thermal Hall phonon drag in a compensated metal. In contrast to metallic SrTiO$_3$~\cite{Jiang2022}, graphite hosts charge carriers of both signs and $\kappa_{xy}$ and $S_{drag}$ reverse sign as a function of temperature. This indicates that momentum exchange between phonons and charge carriers evolves as a function of temperature. Phonon drag is electron-dominated at low temperatures but becomes hole-dominated  at room temperature. Remarkably, at moderate temperatures, when these competing contributions become comparable in magnitude, the thermal Hall effect is expected to completely cancel out, which is exactly what we observe around 100 K (see Supplementary Material~\cite{SM} for details).  Thus, the electric Hall conductivity, which does not change sign is set by charge carrier mobilities shown in Figure~\ref{fig:Comparison}e. On the other hand, the Seebeck coefficient under field is dominated by phonon drag ($S_{drag}$) and its sign is set by the carrier which exchanges most momentum with phonons. This changes at $T\sim$ 100 K. A quantitative account of coupling between electrons and holes, located at distinct high-symmetry points of the Brillouin zone, with acoustic phonons is a task for future \textit{ab initio}  calculations.

The Wiedemann-Franz law establishes a fundamental link between charge and heat transport in metals. Table~\ref{tab:sample_properties} lists the Hall Lorenz number ($L_{xy} = \kappa_{xy} / \sigma_{xy} T$)  for different metals spanning a broad range of electrical Hall conductivities—over four orders of magnitude. Remarkably, graphite exhibits a Hall Lorenz number that is significantly larger than those of all other listed materials.

%In summary, our experimental study of the thermal Hall effect in graphite finds that it exceeds by two orders of magnitude the expected value according to the Wiedemann-Franz law, leading to a record Hall Lorenz number in any metal. Moreover, the signal displays a sign reversal with warming. Both these features can be explained assuming phonon drag of an ambipolar variety. High-mobility quasiparticles generate a large Nernst conductivity, the bipolar flow will superimpose it and cancel the electric Hall conductivity~\cite{Bel2003}. Heat conductivity is dominated by phonons. As a consequence, momentum exchange between the phonon bath and the electron-hole reservoir can generate a significant thermal Hall response. Whose sign is set by the sign of the Peltier (Seebeck) phonon drag. The carrier that exchanges the most momentum with phonons dominates the sign of the effect.

%In summary, we have shown that graphite exhibits a giant thermal Hall effect that severely violates the Wiedemann-Franz law, yielding a record-high Hall Lorenz number among metals and undergoing a sign reversal. These features are explained by a unified scenario of ambipolar phonon drag. High-mobility quasiparticles generate a large Nernst conductivity. Heat conductivity is dominated by phonons. As a consequence, momentum exchange between the phonon bath and the  electron-hole reservoir can generate a significant thermal Hall response. Whose sign is set by the sign of the Peltier (or Seebeck) phonon drag, the carrier that exchanges the most momentum with phonons dominates the sign of the effect.

In summary, we have shown that graphite exhibits an enhanced thermal Hall effect that severely violates the Wiedemann-Franz law, yielding a record Hall Lorenz number undergoing a sign reversal. These features are explained by an ambipolar phonon drag scenario. High-mobility quasiparticles generate a large Nernst conductivity. Heat conductivity, dominated by phonons, acquires a transverse component thanks to momentum exchange between the phonon bath and the electron-hole reservoirs. The sign of the thermal Hall response matches the sign of the phonon drag Seebeck effect, reflecting the type of carrier which exchanges most with phonons.

%The mechanism originates from high-mobility quasiparticles that produce a large Nernst conductivity. In a system where phonons dominate heat conduction, this enables significant momentum transfer between the phonon bath and the electron-hole fluid, thereby generating the observed significant thermal Hall signal. The sign of this signal is determined by the asymmetry in electron-phonon and hole-phonon coupling, as reflected in the phonon-drag Peltier (Seebeck) coefficient, where the carrier species exchanging more momentum with phonons dictates the overall sign.

\textit{Acknowledgments}—This work, part of a Cai Yuanpei Franco-Chinese program (No. 51258NK), was supported by The National Key Research and Development Program of China (Grant No. 2023YFA1609600, 2024YFA1611200 and 2022YFA1403500), the National Science Foundation of China (Grant No. 12304065, 51821005, 12004123, 51861135104 and  11574097), the Fundamental Research Funds for the Central Universities (Grant No. 2019kfyXMBZ071), and the Hubei Provincial Natural Science Foundation ‌(2025AFA072).\\

%\textit{Data availability}—The data that support the findings of this Letter are not publicly available upon publication because it is not technically feasible and/or the cost of preparing, depositing, and hosting the data would be prohibitive within the terms of this research project. The data are available from the authors upon reasonable request.\\

%\textit{Supplementary Material}—The supplementary material includes more details of samples, methods, thermocouple configurations, thermal Hall angle and conductivity, electric and thermoelectric transport data, carrier concentrations and mobilities, comparison of Hall Lorenz number, data reproducibility, which includes Refs.~\cite{Tsuzuku1972, Tokumoto2004,Klein1962,Dresselhaus2002,Zhang2000,Jiang2022,Shiomi2009,Onose2008,Li2017,xu2020finite,zhang2021topological,Zhang2024Abnormally,shiomi2010effect,ding2021quantum}.\\

\noindent
* \verb|lixiaokang@hust.edu.cn|\\
* \verb|zengwei.zhu@hust.edu.cn|\\
* \verb|kamran.behnia@espci.fr|\\

\bibliography{main}

\clearpage
% Add 'S' to the numbering inside the supplement
\renewcommand{\thesection}{S\arabic{section}}
\renewcommand{\thetable}{S\arabic{table}}
\renewcommand{\thefigure}{S\arabic{figure}}
\renewcommand{\theequation}{S\arabic{equation}}
\setcounter{section}{0}
\setcounter{figure}{0}
\setcounter{table}{0}
\setcounter{equation}{0}

{\large\bf Supplementary Material for ``Thermal Hall conductivity of semi-metallic graphite dominated by ambipolar phonon drag''}
{\large\bf by Q. Xiang et al.}

\setcounter{figure}{0}

\section{Samples and methods}
The highly oriented pyrolytic graphite (HOPG) crystals used in this work were cut and cleaved to desired dimensions from the same commercially obtained mother crystal. Measuring the thermal Hall effect in graphite is particularly challenging due to its very high longitudinal thermal conductivity. To overcome this challenge, the samples were prepared to be very thin and moderately wide. Such as sample \#1 (used in the main text) measured 6.0 mm (length) $\times 2.5$ mm (width) $\times 7.3 ~\mu $m (thickness), and sample \#2 (used for the data reproducibility in the Supplementary Material) measured 6.0 mm $\times$ 2.5 mm $\times 12.5 ~\mu $m. The exceptionally small thickness reduces the cross-sectional area, which helps establish a measurable longitudinal temperature gradient, while the moderate width allows detection of a sufficiently large transverse temperature difference signal under a fixed thermal gradient.

All transport experiments were performed in two commercial measurement systems (Quantum Design PPMS and Oxford Teslatron PT) within a stable high-vacuum sample chamber. The voltage was monitored by DC-nanometers (Keithley 2182A) and electric current was driven by a current source (Keithley 6221).  One-heater-four-thermocouples (type E) method was employed to simultaneously measure the longitudinal and transverse thermal gradient. The thermal gradient in the sample was produced through a 4.7 k$\Omega$ chip resistor powered by a current source (Keithley 6221). The thermocouples, the heat-sink, and the heater were connected to samples directly. All contacts on the sample were made using silver paste. 

In the main text, we compare the peak values of $L_0\sigma_{xy}T$ and $\kappa_{xy}$ extracted from their field-dependent curves at each temperature. The rationale for this choice is twofold. First, for $L_0\sigma_{xy}T$, the peak corresponds to the upper bound of the electronic contribution to the transverse thermal conductivity. Second, for $\kappa_{xy}$, using the peak value ensures a consistent comparison with $L_0\sigma_{xy}T$ and provides a unified reference point when analyzing electrical, thermoelectric, and thermal transport signals, as the peak represents the most salient feature of each curve.

\section{Thermocouple configurations}
There are two main configurations in the use of thermocouples:  the differential method (left of Figure~\ref{fig:TD}a) and the single-ended method (right of Figure~\ref{fig:TD}a, used in this work). In the differential setup, a single thermocouple pair directly measures the temperature difference between two points on the sample, requiring only one voltmeter. However, this method carries a short-circuit risk when measuring conductive samples, unless the thermocouple junctions are properly insulated (e.g., with Omega 101). In contrast, the single-ended method avoids this issue by making point contact with the sample at each measurement location. Each junction is read independently by separate voltmeters, eliminating closed-loop current paths through the sample. This configuration functions as direct-contact thermometry, providing true local temperatures. Its main drawback is the need for two voltmeters and two sets of leads to determine a single temperature gradient.

 \section{Thermal Hall angle and conductivity}
\begin{figure*}[ht]
\centering
\includegraphics[width=0.85\linewidth]{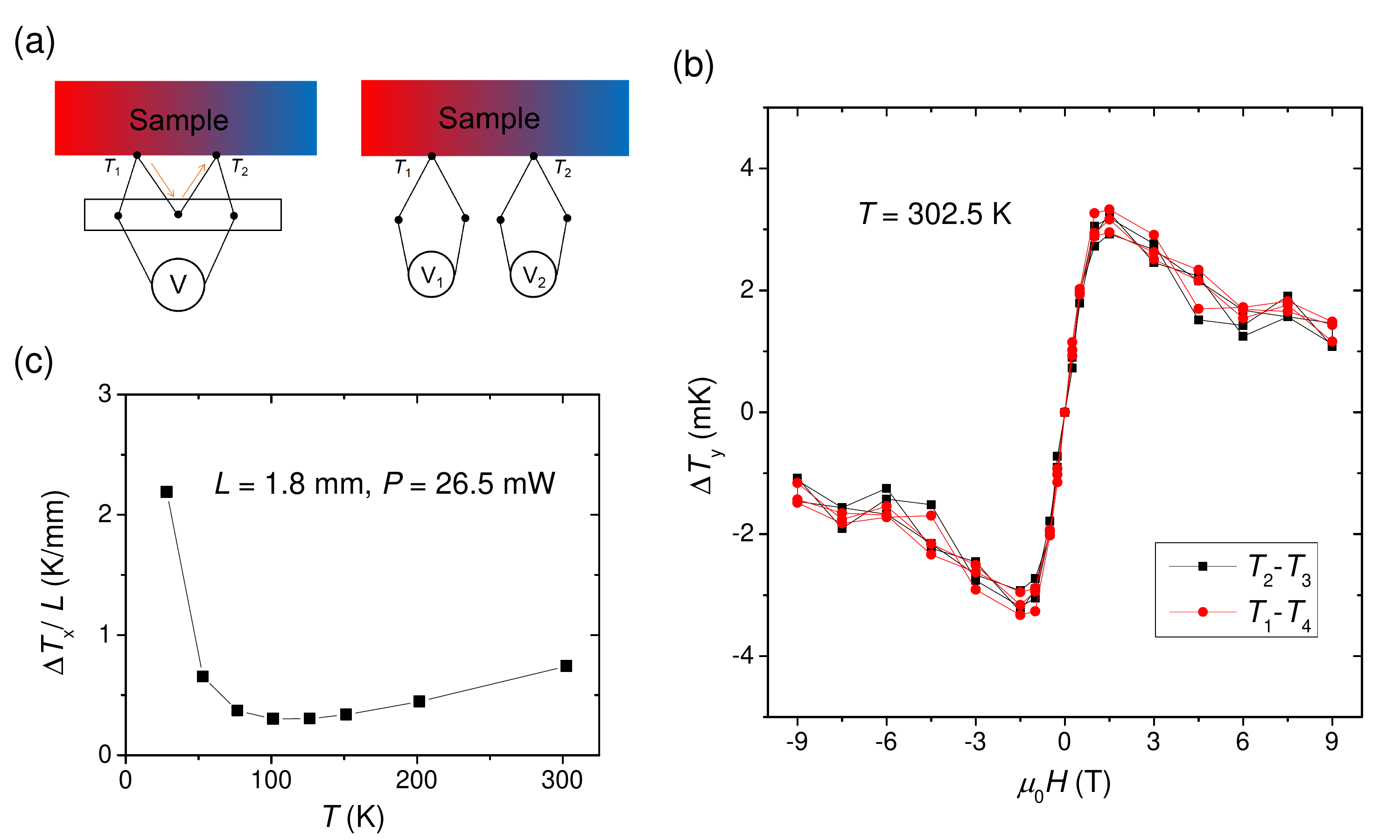} 
\caption{\textbf{Thermocouples and thermal gradient.} Two different thermocouple configurations (a), the longitudinal temperature gradient for the thermal Hall measurements of sample \#1 (b), and two sets of transverse temperature difference at 302.5 K (c).} 
\label{fig:TD}
\end{figure*}

 The one-heater-four-thermocouples (type E) configuration, not only allows simultaneous measurement of both longitudinal ($\nabla T_x = (T_1-T_2)/l$ or $\nabla T_x = (T_4-T_3)/l$) and transverse ($\nabla T_y = (T_2-T_3)/w$ or $\nabla T_y = (T_1-T_4)/w$) thermal gradients induced by a longitudinal thermal current $J_Q$, but also enables the collection and cross-verification of two sets of transverse temperature difference data ($T_2-T_3$ and $T_1-T_4$) to ensure signal reliability and homogeneity, as shown in Figure~\ref{fig:TD}b. For the thermal Hall measurements of sample \#1, the longitudinal temperature gradient ranged from 0.3 to 2.2 K/mm, its temperature dependence is plotted in Figure~\ref{fig:TD}c.
 
 The thermal Hall angle is defined as the ratio of longitudinal to transverse thermal gradients ($\nabla T_y / \nabla T_x$). From the measured thermal gradients and Hall angle, the longitudinal ($\kappa_{xx}$) and the transverse ($\kappa_{xy}$) thermal conductivity can be calculated through:
\begin{equation}\label{kappaii}
\kappa_{xx} = \frac{J_Q}{\nabla T_x}
\end{equation}
\begin{equation}\label{kappaij}
\kappa_{xy} = \frac{\nabla T_y}{\nabla T_x} \cdot \kappa_{xx}
\end{equation}
where $l$ denotes the distance between longitudinal thermocouples, $w$ the sample width, and $J_Q$ the heat power per unit cross-sectional area.  The analysis assumes isotropic in-plane thermal conductivity ($\kappa_{xx} = \kappa_{yy}$). 

\begin{figure*}[ht]
\centering
\includegraphics[width=1\linewidth]{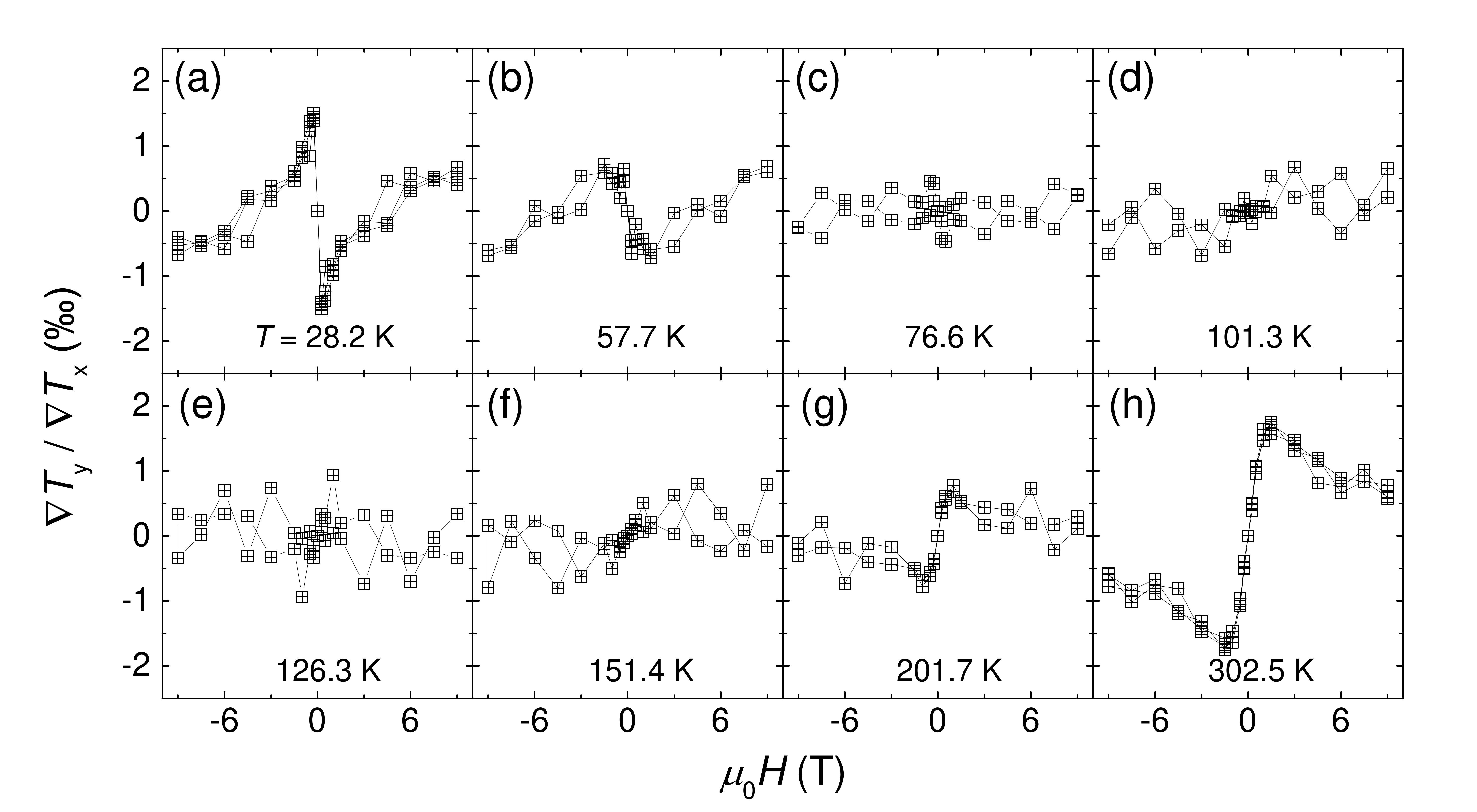} 
\caption{\textbf{Thermal Hall angle.} Field dependence of thermal Hall angle at eight characteristic temperatures ranging from 28.2 K to 302.5 K.
}
\label{fig:THA}
\end{figure*}

\begin{figure*}[ht]
\centering
\includegraphics[width=1\linewidth]{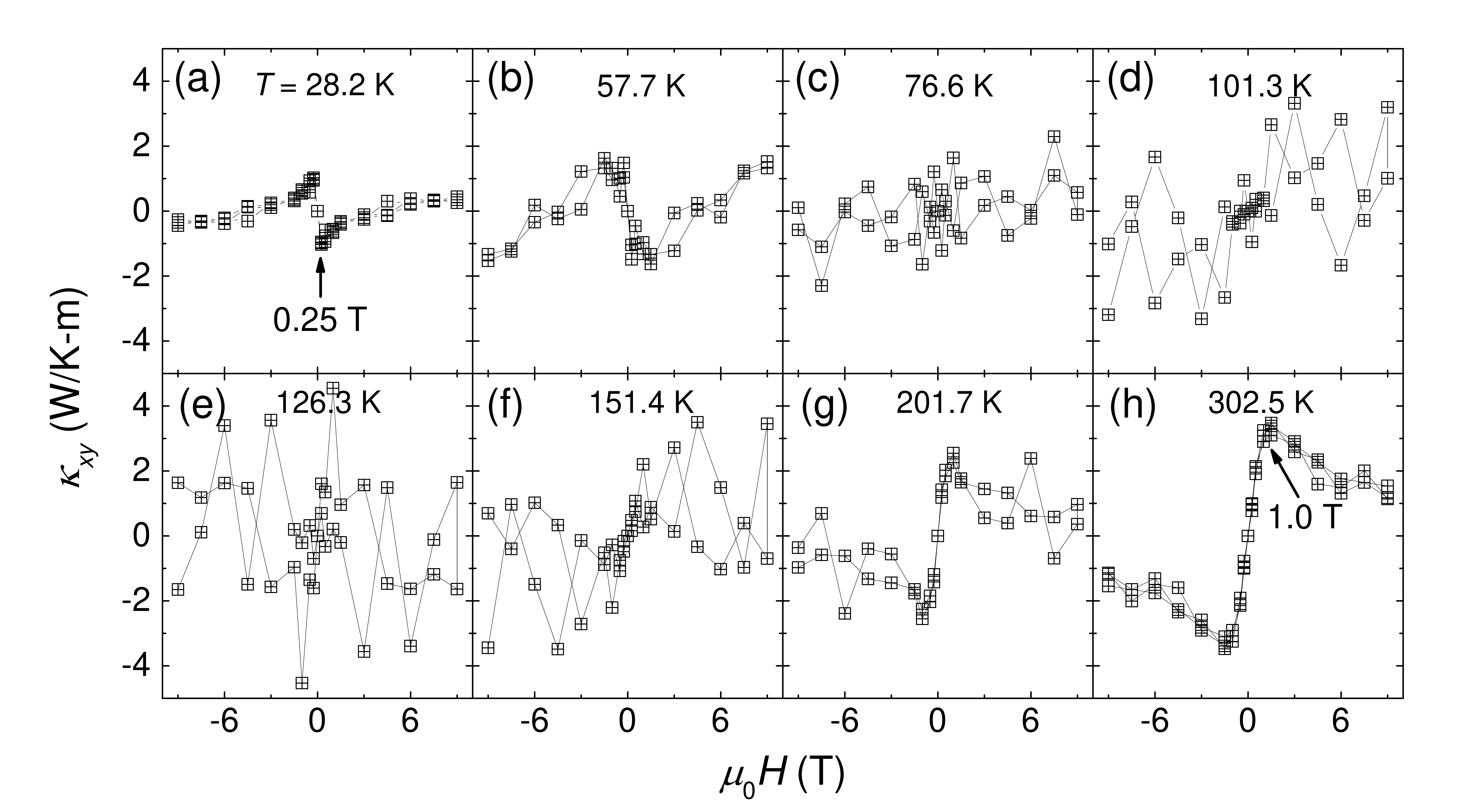} 
\caption{\textbf{Thermal Hall conductivity.} Field dependence of thermal Hall conductivity at eight characteristic temperatures ranging from 28.2 K to 302.5 K.
}
\label{fig:THC}
\end{figure*}

\begin{figure*}[ht]
\centering
\includegraphics[width=0.85\linewidth]{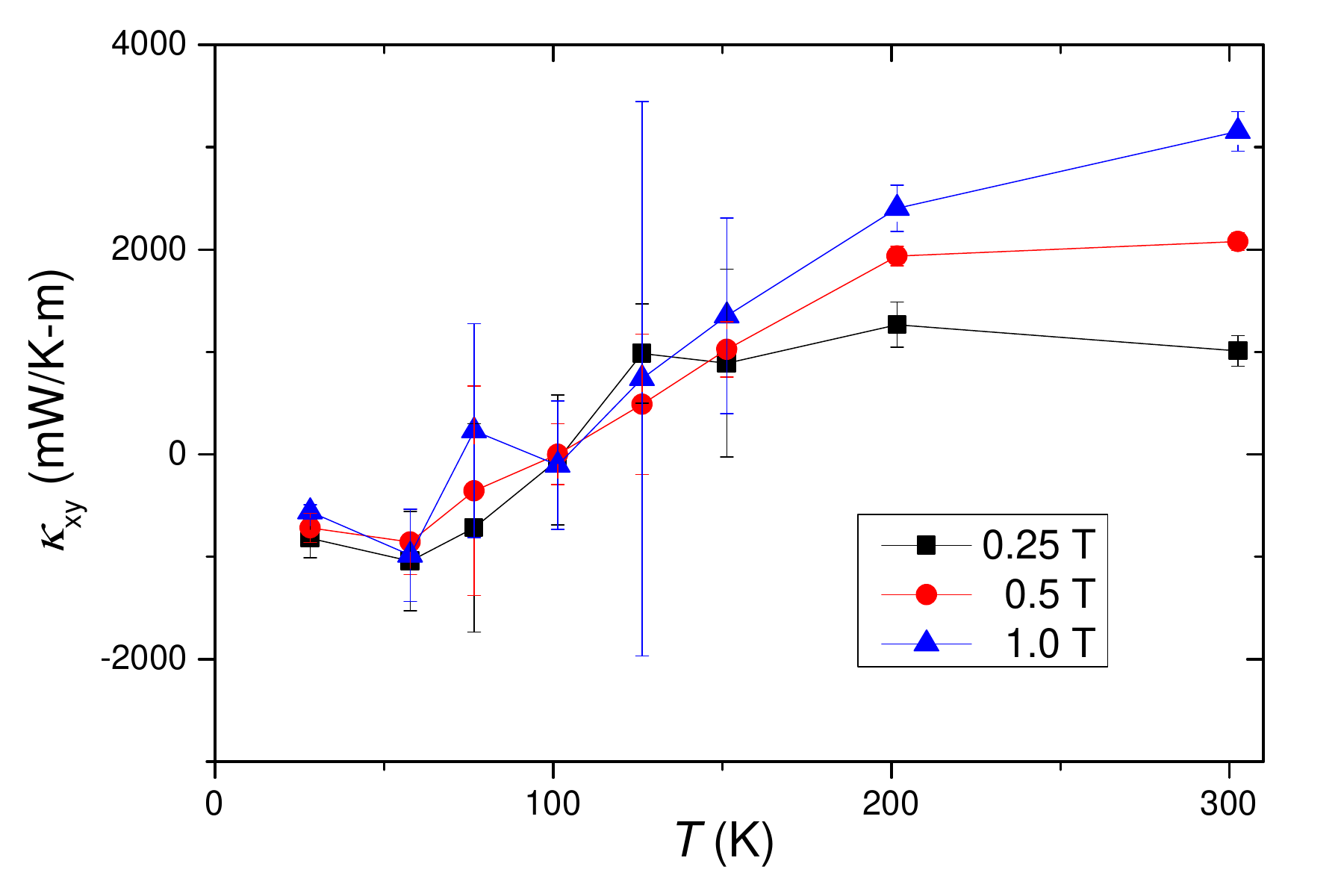} 
\caption{\textbf{Temperature-dependent thermal Hall conductivity under different fields.} 
}
\label{fig:T-THC}
\end{figure*}

Figure~\ref{fig:THA} and Figure~\ref{fig:THC} display the field dependence of $\nabla T_y / \nabla T_x$ and $\kappa_{xy}$ at eight characteristic temperatures ranging from 28.2 K to 302.5 K. A distinct sign reversal emerges near 100 K, which is further evident in the temperature-dependent $\kappa_{xy}$ curves, as shown in Figure~\ref{fig:T-THC}. 

As shown in Figure~\ref{fig:T-THC}, the temperature dependence of $\kappa_{xy}$ at several fixed magnetic fields reproduces the same trend observed in the peak-based analysis in the main text: a negative signal at low temperatures, a sign reversal around 100 K, and a positive signal at high temperatures. This confirms that the peak value reliably captures the intrinsic behavior.

The primary reason for the relatively large noise in the thermal Hall data within the 100–150 K range is the pronounced thermal conductivity peak exhibited by graphite in this interval, which impacts the measurement in several ways. First, it becomes difficult to establish a stable temperature gradient across the sample. Second, when calculating the thermal Hall conductivity $\kappa_{xy}$ (the product of the thermal Hall angle $\nabla T_y / \nabla T_x$ and the longitudinal thermal conductivity $\kappa_{xx}$), the large value of $\kappa_{xx}$ amplifies any noise present in the thermal Hall angle. Furthermore, since this temperature range coincides with the sign‑reversal region of the intrinsic signal, the measured response becomes dominated by experimental noise.

\section{Nernst conductivity}
\begin{figure*}[ht]
\centering
\includegraphics[width=0.9\linewidth]{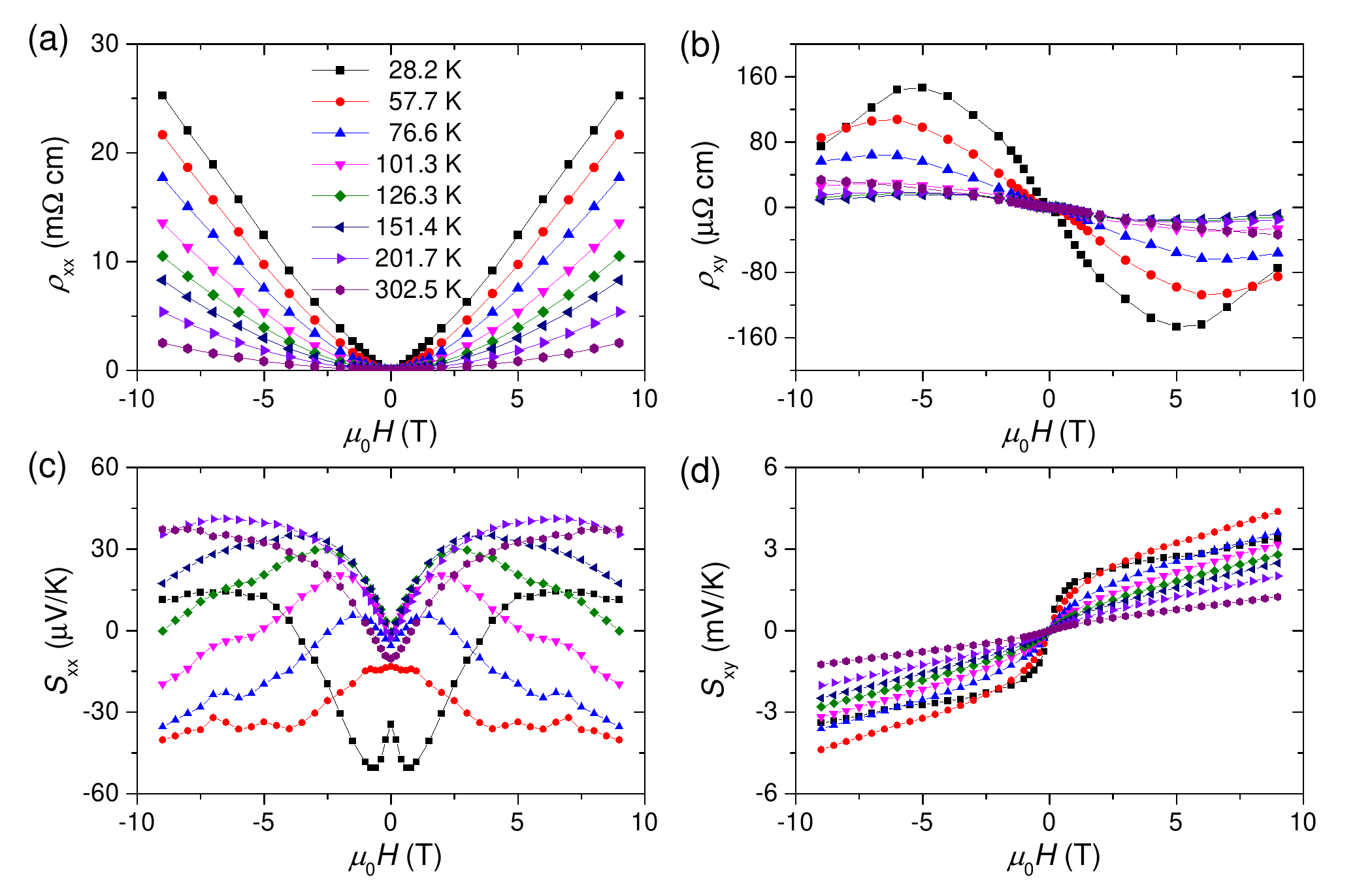} 
\caption{\textbf{Longitudinal and transverse electric/thermoelectric transport signal.} Field dependence of resistivity (a), Hall resistivity (b), Seebeck coefficient (c) and Nernst coefficient (d) at eight characteristic temperatures ranging from 28.2 K to 302.5 K.
}
\label{fig:rho-and-S}
\end{figure*}

\begin{figure*}[ht]
\centering
\includegraphics[width=1\linewidth]{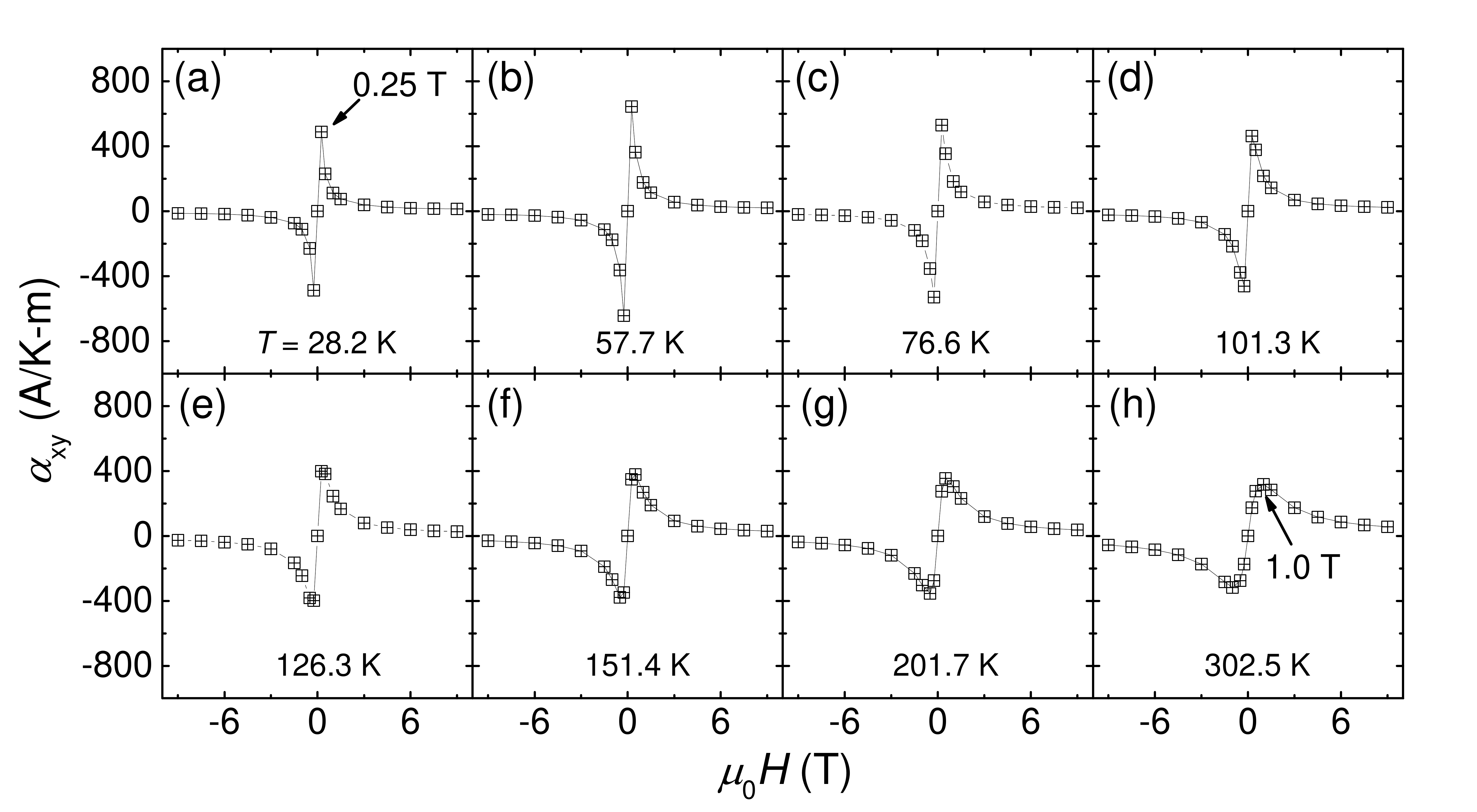} 
\caption{\textbf{Nernst conductivity.} Field dependence of Nernst conductivity at eight characteristic temperatures ranging from 28.2 K to 302.5 K.
}
\label{fig:NC}
\end{figure*}

\begin{figure*}[ht]
\centering
\includegraphics[width=0.85\linewidth]{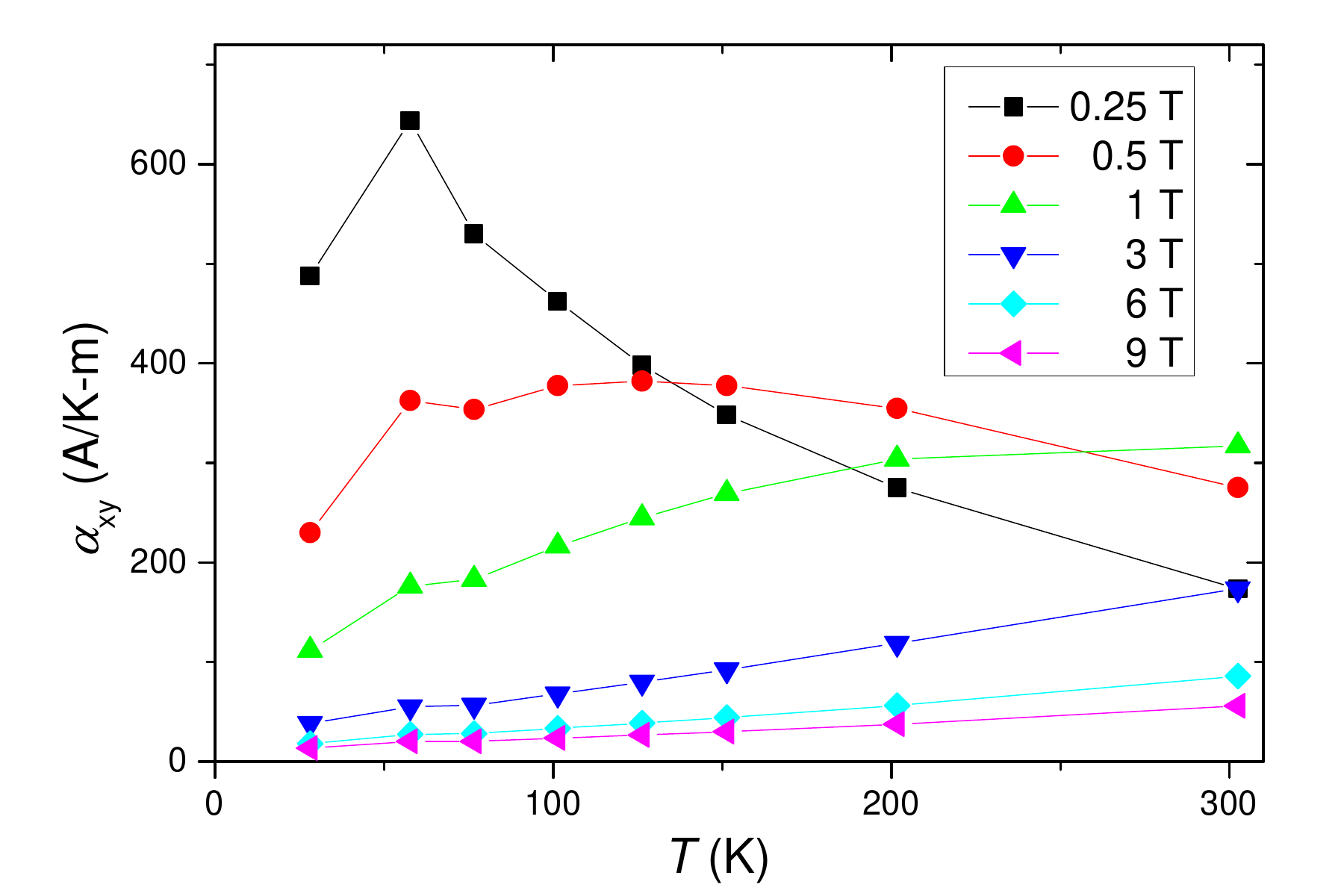} 
\caption{\textbf{Temperature-dependent Nernst conductivity under different fields.} 
}
\label{fig:T-NC}
\end{figure*}

The field-dependent longitudinal and transverse electric/thermoelectric transport data measured at various temperatures are displayed in Figure~\ref{fig:rho-and-S}. The Nernst conductivity (shown in Figure~\ref{fig:NC} and Figure~\ref{fig:T-NC}) is derived from these four transport parameters using the following equation:
\begin{equation}\label{kappaiii}
\alpha_{xy} = \frac{\rho_{xx}S_{xy}-\rho_{xy}S_{xx}}{\rho_{xx}^2}
\end{equation}
Here, $\rho_{xx}$, $\rho_{xy}$, $S_{xx}$, $S_{xy}$ are resistivity, Hall resistivity, Seebeck coefficient and Nernst coefficient respectively. This analysis adopts the isotropic in-plane resistivity assumption ($\rho_{xx} = \rho_{yy}$).

\section{Comparison of Seebeck curves}
\begin{figure*}[ht]
\centering
\includegraphics[width=0.85\linewidth]{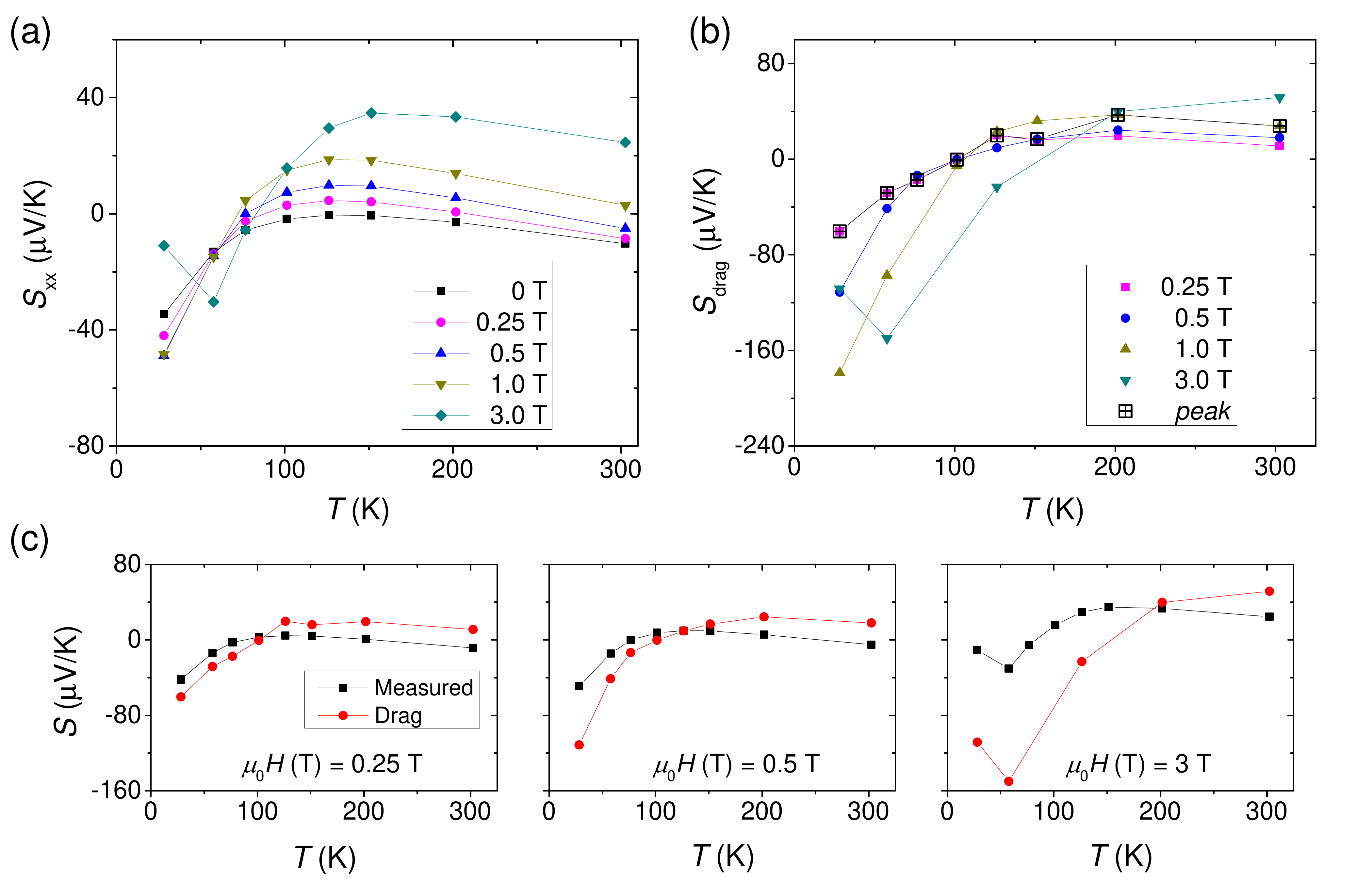} 
\caption{\textbf{Comparison of Seebeck coefficient.} Comparison of the measured Seebeck coefficient $S_{xx}$ (a) with the extracted phonon drag Seebeck coefficient $S_{drag}$ (b) under different fields varying from 0 T to 3 T. The \textit{peak} in (b) denotes the field value corresponding to the maximum of the $\kappa_{xy}$ and $\alpha_{xy}$ curves. The latter agrees better with the former at low field and low temperature, and at high field and high temperature (c).}
\label{fig:T-S}
\end{figure*}

Figure~\ref{fig:T-S} compares the measured Seebeck coefficient $S_{xx}$  with the extracted phonon drag Seebeck coefficient $S_{drag}$ under different field varying from 0 T to 3 T. The latter agrees better with the former at low field and low temperature, and at high field and high temperature.

\begin{figure*}[ht]
\centering
\includegraphics[width=0.85\linewidth]{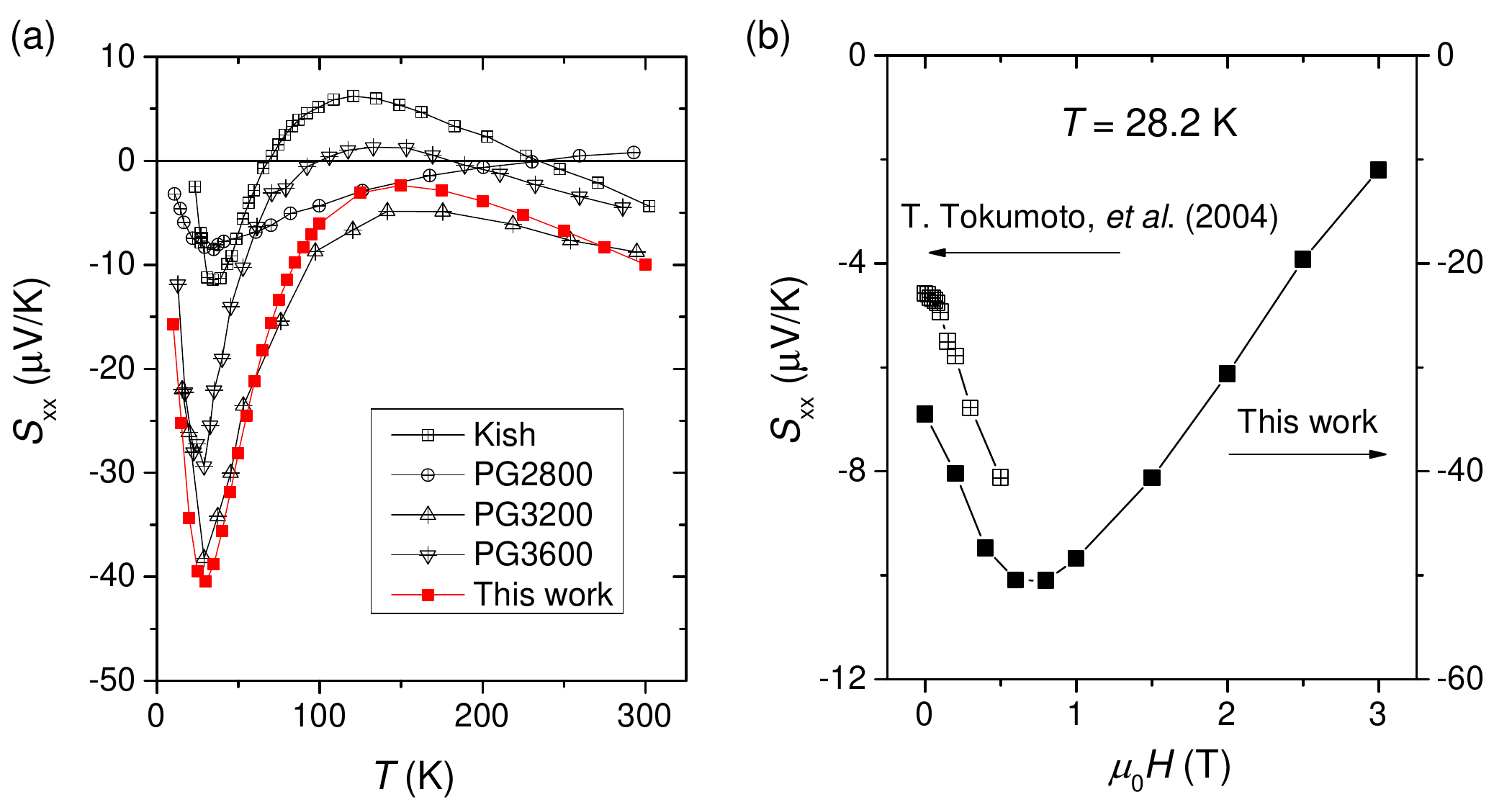} 
\caption{\textbf{Comparison of the Seebeck signal in this work with literature~\cite{Tsuzuku1972, Tokumoto2004}.}} 
\label{fig:C-S}
\end{figure*}

Figure~\ref{fig:C-S} compares the Seebeck signal measured in this study with what was previously reported ~\cite{Tsuzuku1972, Tokumoto2004}. Figure~\ref{fig:C-S}a shows that the samples used in this work are close to the PG3200 sample in the literature~\cite{Tsuzuku1972}.

\section{Carrier concentrations and mobilities of electrons and holes}
\begin{figure*}[ht]
\centering
\includegraphics[width=0.85\linewidth]{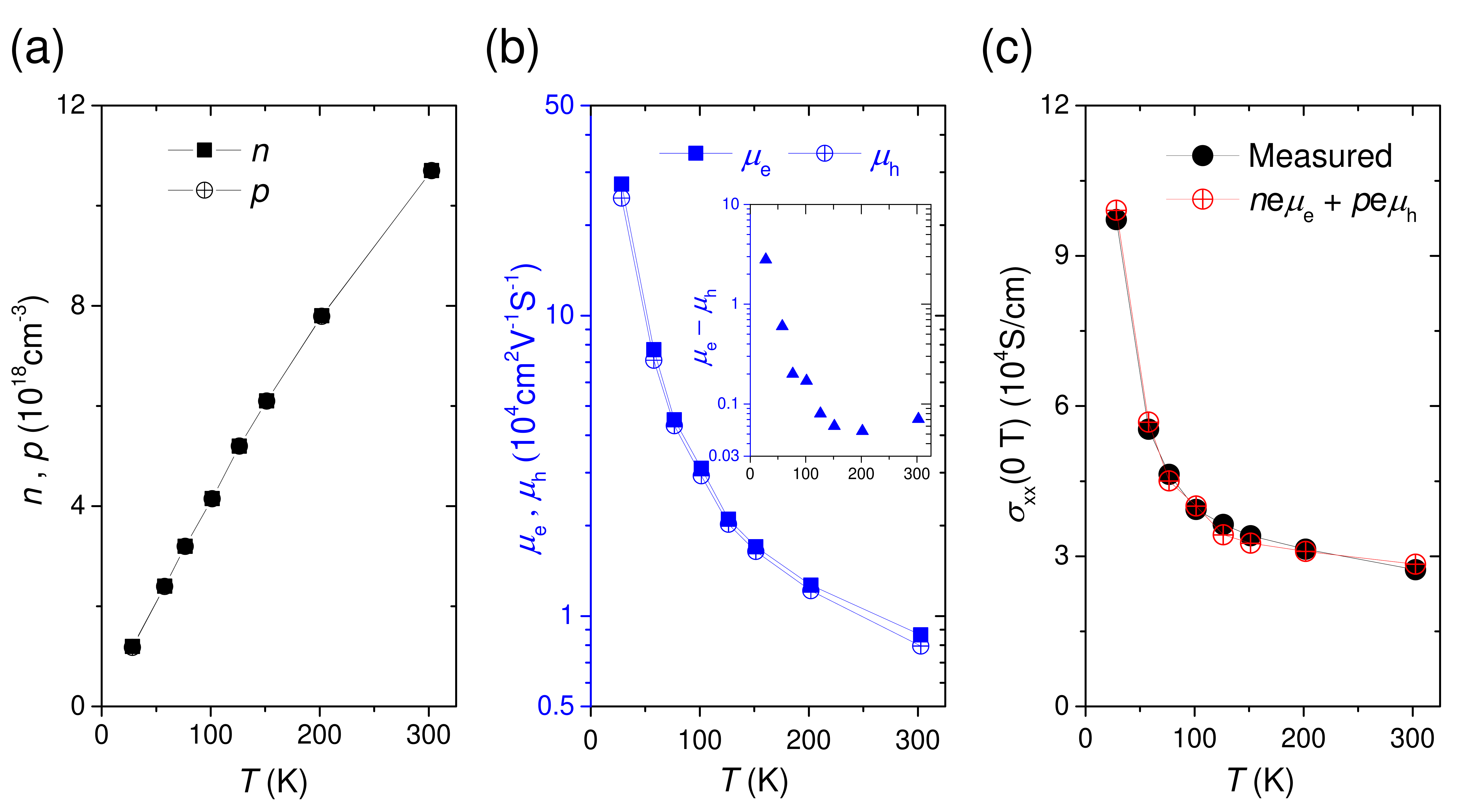} 
\caption{\textbf{Carrier concentrations and mobilities of electrons and holes.} (a-b) Temperature evolution of the carrier concentrations and mobilities of electrons and holes, extracted from the two-band fitting ($\sigma_H(B) = \frac{ne\mu_e^2B}{1+\mu_e^2B^2} - \frac{pe\mu_h^2B}{1+\mu_h^2B^2} $). (c) Comparison between the measured longitudinal electric conductivity ($\sigma_{xx}$) and the estimated value using a two-band model ($ne\mu_e + pe\mu_h $).
}
\label{fig:Carriermobilities}
\end{figure*}

Figure~\ref{fig:Carriermobilities}a-b show the temperature evolution of the carrier concentrations and mobilities of electrons and holes, extracted from the two-band fitting ($\sigma_H(B) = \frac{ne\mu_e^2B}{1+\mu_e^2B^2} - \frac{pe\mu_h^2B}{1+\mu_h^2B^2} $). Figure~\ref{fig:Carriermobilities}c compares the measured longitudinal electric conductivity ($\sigma_{xx}$) and the estimated value using a two-band model ($ne\mu_e + pe\mu_h $). The consistency strengthens our quantification of the temperature dependence of mobilities and carrier densities. 

\begin{figure*}[ht]
\centering
\includegraphics[width=0.85\linewidth]{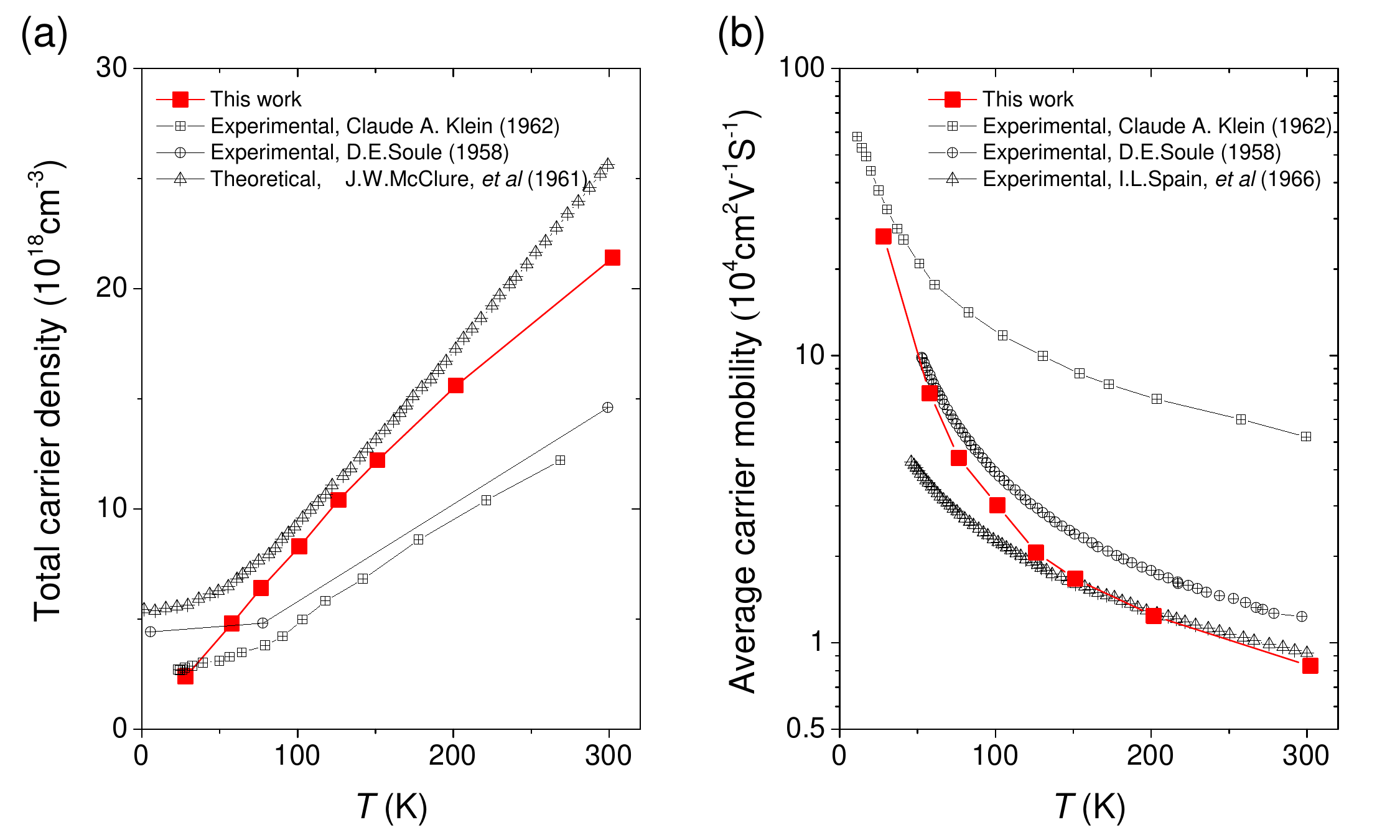} 
\caption{\textbf{Comparing the temperature-induced changes in carrier density (a) and mobility(b) with data from the literature~\cite{Klein1962}.} 
}
\label{fig:Carriermobilities_comparison}
\end{figure*}

Figure~\ref{fig:Carriermobilities_comparison} compares the temperature dependence of carrier density and mobility with previous reports~\cite{Klein1962}, showing results that are consistent with prior works. The temperature-induced change in the carrier density suggests an effective band gap (0.02–0.04 eV) in graphite~\cite{Dresselhaus2002}.

\section{Comparison of Hall Lorenz number in metals}
\begin{figure*}[ht]
\centering
\includegraphics[width=0.85\linewidth]{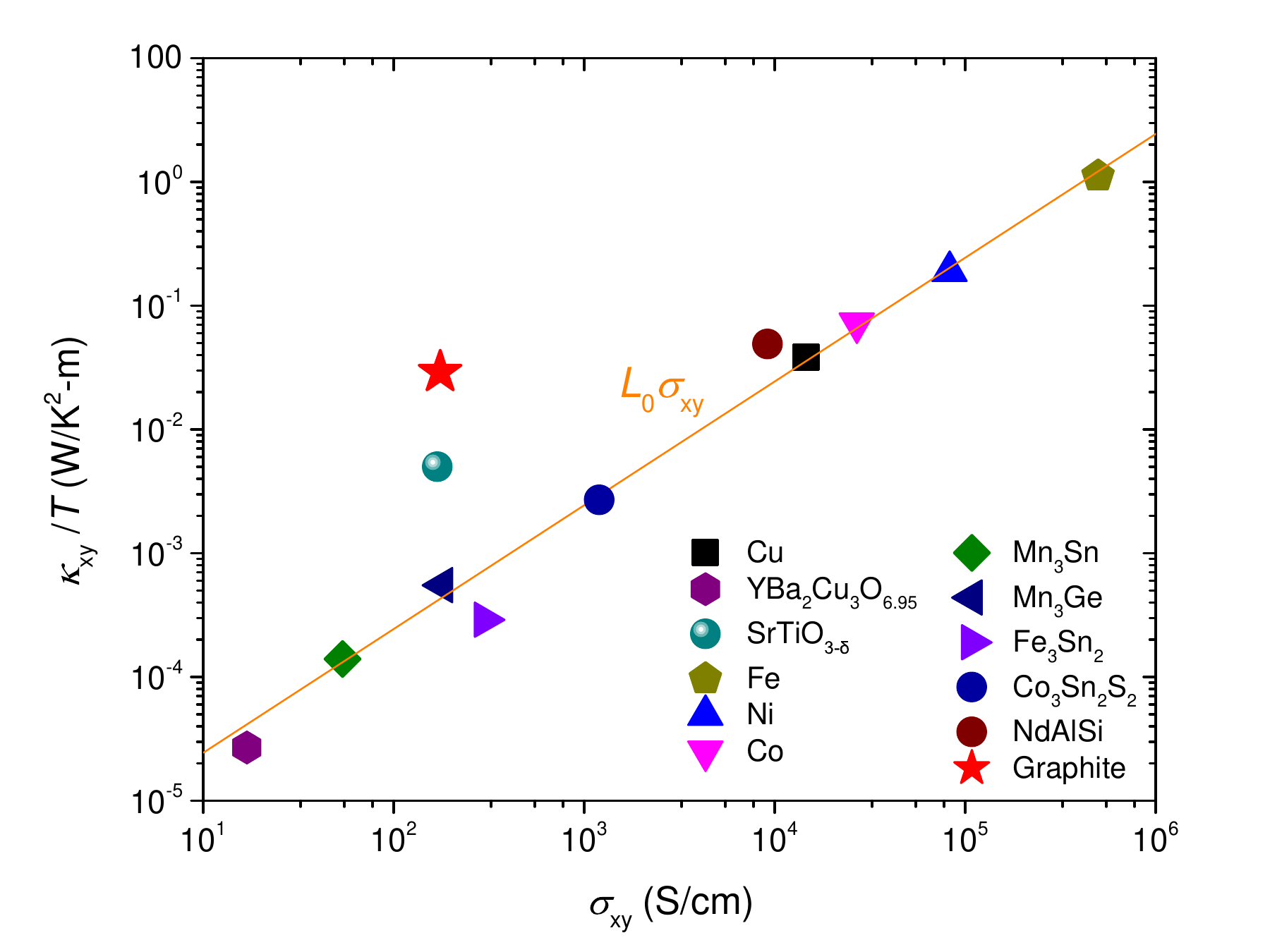} 
\caption{\textbf{The Hall Lorenz number (${L}_{xy}$) in different metals (source:~\cite{Zhang2000,Jiang2022,Shiomi2009,Onose2008,Li2017,xu2020finite,zhang2021topological,Zhang2024Abnormally,shiomi2010effect,ding2021quantum}).} 
}
\label{fig:S-Lxy}
\end{figure*}

Figure~\ref{fig:S-Lxy} shows $\kappa_{xy}/T$ versus $\sigma_{xy}$ for different metals, with electrical Hall conductivities spanning over four orders of magnitude. The orange line denotes the Sommerfeld value $L_0$ ($2.44\times10^{-8}~\mathrm{V}^2~\mathrm{K}^{-2}$). It is clearly seen that graphite exhibits a significantly large Lorenz number among the listed materials.

\section{Data reproducibility}
\begin{figure*}[ht]
\centering
\includegraphics[width=0.85\linewidth]{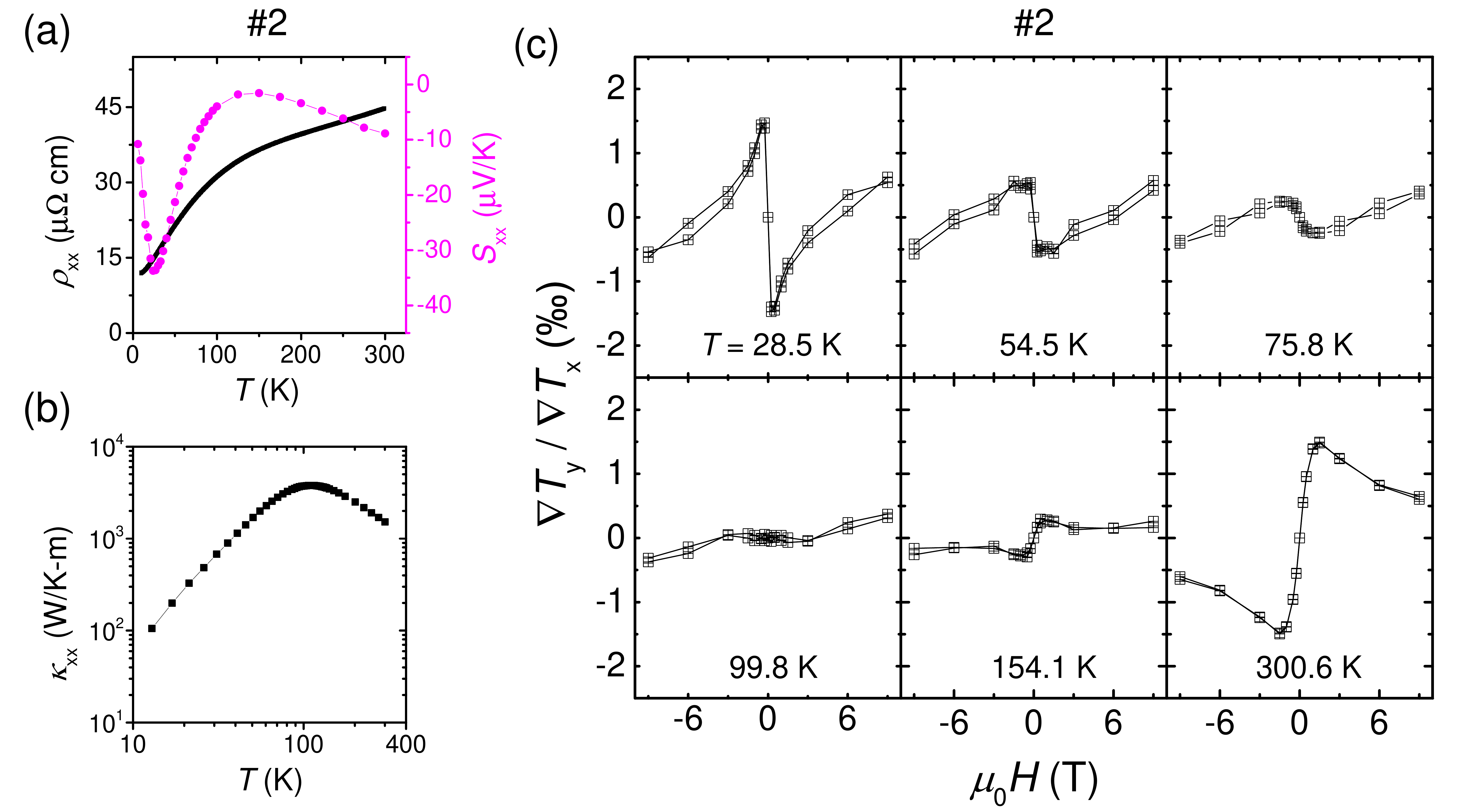} 
\caption{\textbf{Reproducibility of the data in Sample \#2.} (a-b) Temperature dependence of resistivity ($\rho_{xx}$), Seebeck coefficient ($S_{xx}$) and thermal conductivity ($\kappa_{xx}$) in Sample \#2. (c) Field dependence of thermal Hall angle at six characteristic temperatures ranging from 28.5 K to 300.6 K in Sample \#2.
}
\label{fig:S-2}
\end{figure*}

Figure~\ref{fig:S-2} shows the reproducibility of the data in Sample \#2. Figure~\ref{fig:S-2}a and b show the temperature dependence of resistivity ($\rho_{xx}$), Seebeck coefficient ($S_{xx}$) and thermal conductivity ($\kappa_{xx}$) in Sample \#2. Figure~\ref{fig:S-2}c shows the field dependence of thermal Hall angle at six characteristic temperatures ranging from 28.5 K to 300.6 K in Sample \#2. The results measured for Sample \#2 are consistent with those for Sample \#1 used in the main text.

\end{document}